\documentclass[aps,prc,twocolumn,showpacs,superscriptaddress,footinbib]{revtex4-1}

\usepackage{graphicx}
\usepackage{dcolumn}
\usepackage{bm}
\usepackage{enumerate}

\usepackage[normalem]{ulem}  
\usepackage[dvips]{color}

\begin{document}


\title{Fragments' internal and kinetic temperatures in the framework of a Nuclear Statistical Multifragmentation Model}

\author{S.R. Souza}
\affiliation{Instituto de F\'\i sica, Universidade Federal do Rio de Janeiro Cidade Universit\'aria, \\CP 68528, 21941-972, Rio de Janeiro, Brazil}
\affiliation{Instituto de F\'\i sica, Universidade Federal do Rio Grande do Sul,\\
Av. Bento Gon\c calves 9500, CP 15051, 91501-970, Porto Alegre, Brazil}
\author{B.V. Carlson}
\affiliation{Departamento de F\'\i sica, Instituto Tecnol\'ogico de Aeron\'autica-CTA, 12228-900, S\~ao Jos\'e dos Campos, Brazil}
\author{R. Donangelo}
\affiliation{Instituto de F\'\i sica, Universidade Federal do Rio de Janeiro Cidade Universit\'aria, \\CP 68528, 21941-972, Rio de Janeiro, Brazil}
\affiliation{Instituto de F\'\i sica, Facultad de Ingenier\'\i a, Universidad de la Rep\'ublica, Julio Herrera y Reissig 565, 11.300 Montevideo, Uruguay}
\author{W.G. Lynch}
\affiliation{National Superconducting Cyclotron Laboratory and Department of Physics and Astronomy Department,\\ Michigan State University, East Lansing, Michigan 48824, USA}
\author{M.B. Tsang}
\affiliation{National Superconducting Cyclotron Laboratory and Department of Physics and Astronomy Department,\\ Michigan State University, East Lansing, Michigan 48824, USA}

\date{\today}

\begin{abstract}
The agreement between the fragments' internal and kinetic temperatures with the breakup temperature is investigated using a Statistical Multifragmentation Model which makes no a priori assumption on the relationship between them.
We thus examine the conditions for obtaining such agreement and find that, in the framework of our model, this holds only in a relatively narrow range of excitation energy.
The role played by the qualitative shape of the fragments' state densities is also examined. 
Our results suggest that the internal temperature of the light fragments may be affected by this quantity, whose behavior may lead to constant internal temperatures over a wide excitation energy range.
It thus suggests that the nuclear thermometry may provide valuable information on the nuclear state density.
\end{abstract}

\maketitle

\begin{section}{Introduction}
\label{sect:introduction}
The determination of the temperature in the freeze-out configuration of the system, after the most violent stages of a nuclear collision, has challenged both theorists 
\cite {temperaturesBondordf1989,temperaturesBSB1989,temperatureShlomo,temperaturesKonrad1989,PochodzallaReview1997,thermometry2000,thermometry2000,temperaturesTrautmann2007,temperatureReviewMSU1994,BorderiePhaseTransition2008} and experimentalists \cite{reviewSubal2001,thermometryVient2006,BorderiePhaseTransition2008,temperaturesKonrad1989,temperatureReviewMSU1994,reviewTempeatureNatowitz,PochodzallaReview1997,temperaturesHuang1997,thermometry2000,temperaturesTrautmann2007,ccGSI,ccNatowitzHarm,ccMa1997}
 for decades.
The motivation for such studies lies in the theoretical interest in determining the properties of nuclei at finite temperatures and, more generally, of nuclear matter far from the saturation point, {\it i.e.}, its equation of state.
Besides the intrinsic relevance to the research field, such information is of great importance to understand the dynamics of supernovae \cite{BetheEOScollapseStars1979,Bethe1990,EOS2}, where the composition of matter at low densities and finite temperatures, reached during some of their stages, is fairly sensitive to the properties of the hot nuclei \cite{BotvinaSNova2004,BotvinaSNova2005,supernovae2009}.

Despite many efforts, measuring the nuclear temperature turned out to be a rather subtle task and important discrepancies have been found among the different methods that have been employed  \cite{reviewSubal2001,thermometryVient2006,BorderiePhaseTransition2008,reviewTempeatureNatowitz,temperatureReviewMSU1994}.

Indeed, if the energy deposited into the system is very low, it releases most of its excitation energy by a single particle emission, besides electromagnetic transitions, so that the energy spectra of particles emitted from such systems should provide information on the temperature of the hot source, named kinetic temperature and labelled $T_{\rm kin}$ henceforth.
On the other hand, as the excitation energy increases, more and more particles are emitted during the decay chain and the final energy spectra are influenced by the properties of the system at the different stages.
Furthermore, distortions due to the recoil of the source, fluctuations of the Coulomb barrier, besides those associated with the decay of the emitted particles, should also be taken into account  \cite{temperatureReviewMSU1994}.
In practice, the temperatures derived from the energy spectra of fragments originated from multiparticle emission are higher than expected  \cite{temperaturesBondordf1989,kineticTemperaturesBauer1995} and may vary according to the selected species \cite{temperatureReviewMSU1994,temperaturesKonrad1989,temperaturesBetty1996}.

Based on the isotopic abundances of fragments produced in the reaction, Albergo {\it et al.} \cite{Albergo} derived an expression for the isotopic temperature $T_{\rm iso}$, which has been widely used by different groups.
Upon assuming that chemical and thermal equilibrium are reached at the same time, they obtained a simple expression relating the binding energies of the selected fragments $\{B(A,Z)\}$ and their corresponding yields $\{Y(A,Z)\}$, where $A$ and $Z$ denote the mass and atomic numbers, respectively:

\begin{equation}
\frac{Y(A_1,Z_1)/Y(A_1+1,Z_1)}{Y(A_2,Z_2)/Y(A_2+1,Z_2)}=C\exp(\Delta B/T_{\rm iso})\;,
\label{eq:Albergo}
\end{equation}

\noindent
where $\Delta B\equiv B(A_1,Z_1)-B(A_1+1,Z_1)-B(A_2,Z_2)+B(A_2+1,Z_2)$ and $C$ is a factor associated with the ground state spins of the fragments and kinematic factors.
The fragments are selected in such a way that $\Delta B$ is maximized, in order to minimize the uncertainties of $T_{\rm iso}$.
However, since the fragments are produced in excited states, the yields are distorted by their own decay, besides side feeding due to the deexcitation of heavier fragments.
Therefore, the temperatures obtained from the primary fragments differ, in general, from those calculated with the final yields \cite{thermometry2000,temperatureSecDecRaduta,temperatureCalibrationRaduta1999,thermoetryHeliumMSU1998,secDecIsotopeRatioMSU1999}.
Furthermore, it has also turned out that the treatment employed to describe the deexcitation of the hot primary fragments affects the final isotopic temperatures \cite{thermometry2000,secDecIstoTempModelMSU1996}.
All those aspects contribute to the qualitative differences observed in the experimental caloric curves reported in the recent literature \cite{reviewTempeatureNatowitz,ccNatowitzHarm,ccGSI,ccMa1997,BorderiePhaseTransition2008,thermometryVient2006,temperaturesTrautmann2007} obtained using $T_{\rm iso}$.

The temperature $T_{\rm em}$ at which a species is emitted has been extensively studied experimentally \cite{temperatureReviewMSU1994,reviewTempeatureNatowitz,temperaturesTrautmann2007,BorderiePhaseTransition2008,ccSerfling,ccXi,temperaturesHuang1997,thermoetryHeliumMSU1998,temperatureReviewMSU1994} under the assumption that the  states of the species are populated statistically.
More specifically, in this case, the ratio between the yields of two bound states $a$ and $b$ of the species $(A,Z)$ reads:

\begin{equation}
\frac{Y_a }{Y_b} = \frac{2j_a+1}{2j_b+1}\exp\left[-\left(E_a-E_b\right)/T_{\rm em}\right]\;,
\label{eq:tem}
\end{equation}

\noindent
where $j_a$ ($j_b$) denotes the spin of state $a$ ($b$) and $E_a$ ($E_b$) represents the fragment's excitation energy associated with this state.
As discussed, for instance, in Ref.\ \cite{temperatureReviewMSU1994}, the efficiency of this method is seriously impacted by the saturation of this expression for temperatures larger than $|E_a-E_b|$.
Furthermore, the deexcitation of heavier fragments increases the ground state yields of the considered species, which, from Eq.\ (\ref{eq:tem}), mimics lower emission temperatures \cite{temperatureReviewMSU1994,particleStableStatesMSU}.
Although this effect is appreciably reduced if one avoids the ground state in the above expression, the comparison with theoretical calculations becomes very difficult as very detailed information on the deexcitation process is needed, which quickly becomes prohibitively laborious as the fragment's size increases.
Particle unstable states have also been used in such studies \cite{breakDownStatisticalEquilibriumInternalTemperatures,temperatureReviewMSU1994,ccSerfling,ccXi,impactParameterExcitedStatePopulations,temperaturesTrautmann2007} as it allows the use of larger energy differences than in the case of bound states, so that the limitation to low values of $T_{\rm em}$ should be eliminated.
Nevertheless, although $T_{\rm iso}$ was found to increase as a function of the bombarding energy, $T_{\rm em}$ turned out to saturate at around 5--6 MeV \cite{temperatureReviewMSU1994,temperaturesTrautmann2007,temperaturesHuang1997,ccXi,ccSerfling}, which may be an indication that the limiting temperature predicted by Hartree-Fock calculations \cite{BLV1,BLV2} is attained by the fragments during the multifragment emission, but further studies are needed to draw precise conclusions on this aspect.

Statistical models have been widely employed in the study of the nuclear thermometry \cite{ISMMlong,nuclearThermometry2000,temperatureSecDecRaduta,temperatureCalibrationRaduta1999,isotopicTemperaturesBotvina1998}.
Particularly, the Statistical Multifragmentation Model (SMM) \cite{smm1,smm2,smm4} has been modified in Ref.\ \cite{temperaturesBSB1989} in order to decouple the kinetic and internal temperatures.
More specifically,  $T_{\rm kin}$ was treated as an input parameter, so that the entropy was maximized while keeping $T_{\rm kin}$ fixed.
It was found in that study that the internal temperatures are not very sensitive to $T_{\rm kin}$ for a relatively wide range of values.

In this work we use a modified version of SMM \cite{smmde2013}, in which there is no a priori assumption on the relationship between the internal and kinetic temperatures.
We thus aim at investigating the extent to which the different temperatures agree with each other without imposing any constraints on them.
The influence of the density of states, which enters directly in this formulation of the model, on the different temperatures is also studied.
We start in sec.\ \ref{sect:model} by reviewing the main features of the model, whereas the results are presented and discussed in sec.\ \ref{sect:results}.
The main conclusions are drawn in sec.\ \ref{sect:conclusions}.
 
 \end{section}
 
\begin{section}{Underlying assumptions}
\label{sect:model}
The SMM is based on a scenario \cite{smm1} in which a system of mass and atomic numbers $A_0$ and $Z_0$, respectively, undergoes a prompt breakup at volume $V=(1+\chi)V_0$, where $V_0$ is the volume associated with the normal nuclear density and
$\chi$ is a parameter, whose value is fixed $\chi=2$ through out this work.
The multiplicity of a species $(A,Z)$, with excitation energy $\epsilon^*_{A,Z}$ and kinetic energy $K_{A,Z}$, is represented by $n_{A,Z,q}$, so that

\begin{equation}
q\Delta_Q=-B^c_{A,Z}+\epsilon^*_{A,Z}+K_{A,Z}
\label{eq:q}
\end{equation}

\noindent
stands for its total energy, besides the Coulomb contribution from the Wigner-Seitz calculation of the Coulomb repulsion among the fragments \cite{smm1,smmde2013,WignerSeitz} which, in the above expression, is grouped with the binding energy of the fragment $B_{A,Z}$:

\begin{equation}
B^c_{A,Z}\equiv B_{A,Z}+C_c\frac{Z^2}{A^{1/3}}\frac{1}{(1+\chi)^{1/3}}\;,
\label{eq:bc}
\end{equation}

\noindent
where  $C_c$ is one of the parameters of the mass formula used in the model \cite{smmde2013}, from which  $B_{A,Z}$ is calculated.
The numerical values of the parameters are listed in Ref.\ \cite{ISMMmass}.
The total energy is an input parameter of the model, whose conservation is imposed for each partition:

\begin{equation}
E^*-B^c_{A_0,Z_0}=\Delta_Q \sum_{A,Z,q}q n_{A,Z,q}\;,
\label{eq:econs}
\end{equation} 

\noindent
where $E^*$ denotes the total excitation energy of the source.
The energy is discretized so that $q$ is an integer and $\Delta_Q$ regulates the granularity of the discretization.
The sum in Eq.\ (\ref{eq:econs}) is carried out under the constraint of mass, charge, and energy conservation.

We also introduce a parameter $Q$, which is an integer and is related to the total energy of the source by $Q\Delta_Q=E^*-B^c_{A_0,Z_0}$.
In this way, it is treated as a conserved quantity, on the same footing as $A_0$ and $Z_0$.
It then allows one to employ the efficient recursion relations developed in Ref.\ \cite{PrattDasGupta2000}, from which the average primary multiplicity of a species $(a,z)$ with energy $q\Delta_Q$ may be easily obtained \cite{smmde2013,PrattDasGupta2000}:

\begin{equation}
\overline{n}_{a,z,q}=\frac{\omega_{a,z,q}}{\Omega_{A_0,Z_0,Q}}\Omega_{A_0-a,Z_0-z,Q-q}\;.
\label{eq:nazq}
\end{equation}

\noindent
The recurrence relation \cite{smmde2013,PrattDasGupta2000}

\begin{equation}
\Omega_{A,Z,Q}=\sum_{a,z,q}\frac{a}{A}\omega_{a,z,q}\Omega_{A-a,Z-z,Q-q}\
\label{eq:oazq}
\end{equation}

\noindent
allows one to quickly evaluate the statistical weight $\Omega_{A,Z,q}$ after

\begin{equation}
w_{A,Z,q}=\gamma_A\int_0^{\epsilon_{A,Z,q}}\,dK\;\sqrt{K}\rho_{A,Z}(\epsilon_{A,Z,q}-K)
\label{eq:wazq}
\end{equation}

\noindent
has been calculated.
The parameters $\epsilon_{A,Z}$ and $\gamma_A$ read

\begin{equation}
\epsilon_{A,Z,q}\equiv {\,q\Delta_Q+B^c_{A,Z}}
\label{eq:efrag} 
\end{equation}

\noindent
and

\begin{equation}
\gamma_A\equiv \Delta_Q \frac{V_f (2m_n A)^{3/2}}{4\pi^2\hbar^3}\;.
\label{eq:gamma}
\end{equation}

\noindent
The free volume is denoted by $V_f=\chi V_0$, $m_n$ is the nucleon mass, and $\rho_{A,Z}(\epsilon^*)$ is the density of the internal states of the nucleus $(A,Z)$ with excitation energy $\epsilon^*$.

From the above expressions, it is clear that the state density $\rho_{A,Z}(\epsilon^*)$ plays a key role in determining the statistical properties of the disassembling system.
There are different choices in the literature  \cite{levelDensityGilbertCameron1965,pairingGoriely1996,levelDensityGoriely2001,ldBucurescu2005,ldBucurescu2005Erratum,levelDensityRauscher1997,levelDensityBurtsch2003}, whose complexity and accuracy vary from one parameterization to the other.
In Ref.\ \cite{ISMMlong}, it has been shown that the Helmholtz free energy of the standard SMM can be fairly well described, over a wide range of temperatures, if one adopts the following state density:

\begin{equation}
\rho_{\rm SMM}(\epsilon^*)=\rho_{\rm FG}(\epsilon^*)e^{-b_{\rm SMM}(a_{\rm SMM}\epsilon^*)^{3/2}}
\label{eq:rhoSMM}
\end{equation}

\noindent
with

\begin{equation}
\rho_{\rm FG}(\epsilon^*)=\frac{a_{\rm SMM}}{\sqrt{4\pi}{(a_{\rm SMM}\epsilon^*)}^{3/4}}\exp(2\sqrt{a_{\rm SMM}\epsilon^*})
\label{eq:rhofg}
\end{equation}

\noindent
and

\begin{equation}
a_{\rm SMM}=\frac{A}{\epsilon_0}+\frac{5}{2}\beta_0\frac{A^{2/3}}{T_c^2}\;.
\label{eq:asmm}
\end{equation}

\noindent
This is employed for all nuclei whose $A>4$ and for the alpha particles, in which case the surface term in the above expression is suppressed.
We refer the reader to Ref.\ \cite{ISMMlong} for more details and numerical values for the parameters $\epsilon_0$, $\beta_0$, and $T_c$.
Since we focus on qualitative aspects of the state density, we choose this simple parameterization, and remove the contribution from pairing at high excitation energies by using the back-shifted description \cite{levelDensityGilbertCameron1965}, {\it i.e.} $\rho_{\rm SMM}(\epsilon^*)\rightarrow \rho_{\rm SMM} (\epsilon^*-\Delta)$, where $\Delta$ is the pairing energy.
In order to avoid the divergency at low excitation energy, where the Fermi-gas approximation breaks down, we also proceed as in Ref.\ \cite{levelDensityGilbertCameron1965} and match Eq.\ (\ref{eq:rhoSMM}) with a constant temperature expression, {\it i.e.}

\begin{equation}
\label{eq:rhogc}
\rho(\epsilon^*)=
\cases{
\frac{1}{\tau}e^{(\epsilon^*-E_0)/\tau},&$\epsilon^* \le E_x$,\cr
\rho_{\rm SMM}(\epsilon^*-\Delta),& $\epsilon^* \ge E_x$.
}
\end{equation}

\noindent
The parameters $E_0$ and $\tau$ are determined from the continuity of the state density and its derivative at $\epsilon^*=E_x\equiv U_x+\Delta$.
The parameter $U_x$ has been calculated for many nuclei in Ref.\ \cite{levelDensityGilbertCameron1965} and a parameterization of its average value $U_x=2.5+150.0/A$ MeV has been obtained, for relatively large values of $A$.
We, nevertheless, use this same formula for light nuclei, which might lead to very large values of $U_x$. 
In such cases, the use of discrete states observed experimentally should be more adequate.
However, this is feasible only for very light nuclei, due to the fast increase of the number of states with the nucleus size.
Therefore, one would be forced to adopt some other treatment, such as the use of Eq.\ (\ref{eq:rhogc}), for the heavier nuclei.
Switching from one treatment to the other would lead to peculiar predictions for many observables.
We therefore adopt the simple expression above for all excitable nuclei and, in order to investigate the sensitivity of the results to the behaviour of the state density at low excitation energies, we also use $\tilde U_x={\rm Min}(2.5+150.0/A,5.0)$ MeV.
There is no particular reason for the chosen upper bound, so that slightly different values could have been adopted as well.
We simply aim at investigating the role played by the qualitative properties of the density of states, in the case of light nuclei.

\begin{figure}[tbh]
\includegraphics[width=8.5cm,angle=0]{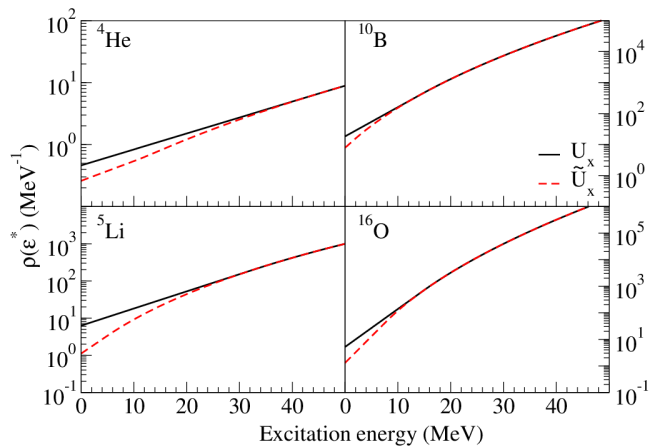}
\caption{\label{fig:rho} (Color online) Density of states for different nuclei. For details see the text.}
\end{figure}

The comparison of the density of states obtained with these two parameters $U_x$ and $\tilde U_x$ is displayed in Fig.\ \ref{fig:rho} for some selected light nuclei, as a function of the total excitation energy.
As expected, it shows that $U_x$ affects the state density, essentially at low excitation energies, and that its influence
diminishes as the system size increases.
It also turns out that, the smaller values of $E_x$, due to the upper bound in $\tilde U_x$, lead to state densities with larger curvatures.
This is due to the fact that, except near $\epsilon^*\rightarrow 0$, $\frac{1}{\tau}e^{-(\epsilon^*-E_0)/\tau}$ rises faster than $\rho_{\rm SMM}(\epsilon^*)$ as $\epsilon^*$ increases.
This effect is particularly noticeable in the case of the $^5$Li nucleus, where the use of $U_x=2.5+150.0/A$ MeV leads to an almost linear density of states in a semilogarithmic plot.
In the case of the alpha particle this behavior is much more pronounced, but it is mainly caused by the parameter set for $\rho_{\rm SMM}(\epsilon^*)$ employed in this case, as both prescriptions for $U_x$ lead to state densities with similar qualitative behaviour.
Since the primary fragments' excitation energy distribution moves to higher values as their size increases \cite{staggering2014}, the influence of the upper bound in $\tilde U_x$ should decrease accordingly.

\end{section}

\begin{section}{Results}
\label{sect:results}
The model described in the previous section is now applied to the breakup of the $^{40}$Ca, $^{80}$Zr, and $^{120}$Nd sources at different excitation energies.
The caloric curve associated with the breakup temperature of these sources, is obtained from:

\begin{equation}
\frac{1}{T}=\frac{\partial\ln(\Omega_{A_0,Z_0,Q})}{\partial (Q\Delta_Q)}\approx \frac{\ln(\Omega_{A_0,Z_0,Q})-\ln(\Omega_{A_0,Z_0,Q-1})}{\Delta_Q}\;.
\label{eq:tbk}
\end{equation}

\noindent
The results obtained with $U_x$ are displayed in the upper left panel of Fig.\ \ref{fig:tkin}.
Since the upper bound in $\tilde U_x$ has a very small influence on $T$, the corresponding results are not exhibited.
In qualitative agreement with previous results \cite{ccIsospinSizeBotvina2002,ccNatowitzHarm}, there is a slight size dependence on the caloric curve and lower temperature values are found for larger system sizes.
The asymptotic limit has apparently been attained around $A_0 \lesssim 80$.

\begin{figure}[tbh]
\includegraphics[width=8.5cm,angle=0]{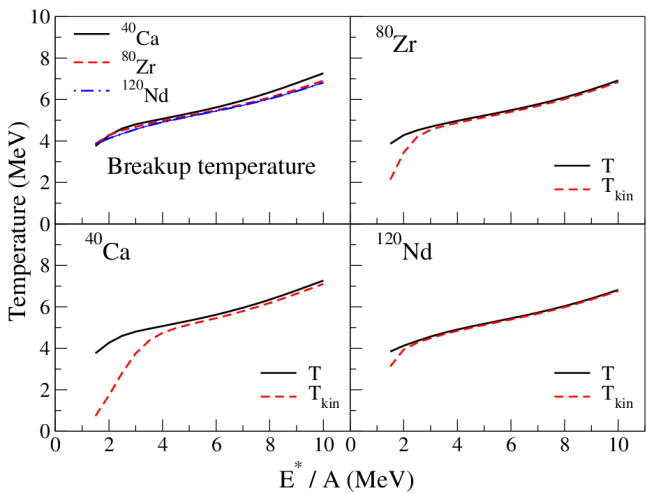}
\caption{\label{fig:tkin} (Color online) Breakup temperature $T$, calculated from Eq.\ (\ref{eq:tbk}), and kinetic temperature $T_{\rm kin}=2 \langle K\rangle / 3$ for different fragmenting sources. For details see the text.}
\end{figure}

The average kinetic energy of the fragments may be obtained with the help of Eqs.\ (\ref{eq:q}), (\ref{eq:nazq}), and (\ref{eq:efrag}):

\begin{equation}
\langle K\rangle=\frac{\sum_{a,z,q} \overline{n}_{a,z,q}\left(\epsilon_{a,z,q}-\overline{\epsilon}^*_{a,z,q}\right)}{\sum_{a,z,q} \overline{n}_{a,z,q}}\;,
\label{eq:aveK}
\end{equation}

\noindent
where $\overline{\epsilon}^*_{a,z,q}$ is the average excitation energy of the species associated with its total energy $\epsilon_{a,z,q}$ \cite{smmde2013}:

\begin{equation}
\overline{\epsilon}^*_{a,z,q}=\frac{\gamma_a}{\omega_{a,z,q}}\int_0^{\epsilon_{a,z,q}}\,dK\;(\epsilon_{a,z,q}-K)\sqrt{K}\rho(\epsilon_{a,z,q}-K)\;.
\label{eq:eex}
\end{equation}

\noindent
In the framework of this model, the kinetic temperature is defined as $T_{\rm kin}\equiv 2 \langle K\rangle / 3$.
A comparison of $T_{\rm kin}$ with the breakup temperature $T$ is also shown in Fig.\ \ref{fig:tkin}, for the three sources studied in this work.
The results clearly reveal discrepancies between the two temperatures at low excitation energies.
The agreement quickly improves as the source's size increases, being very good for $A_0\gtrsim 80$, which suggests that finite size effects
rapidly become negligible in this context.
We have checked that the discrepancies at low excitation energies are reduced if only light fragments, {\it i.e.} those with $Z\le 15$, are included. 
These findings contrast with the simplification made in the standard SMM \cite{smm1,smm2} in which both temperatures are assumed to be the same.
Nevertheless, since the differences are appreciable only for light sources and at low excitation energies, our results give support to this simplifying assumption.
We have also checked that the results are not significantly affected by the adoption of a slightly different density of states, by replacing $U_x$ by $\tilde U_x$ in Eq. (\ref{eq:rhogc}).

We now turn to the internal fragments' temperature which we define as:

\begin{equation}
\frac{1}{T_{\rm int}^{a,z,q}}=\frac{\partial\ln[\rho_{a,z}(\epsilon^*)]}{\partial\epsilon^*}\mid_{\epsilon^*=\overline{\epsilon}^*_{a,z,q}}\;,
\label{eq:tint}
\end{equation}

\noindent
so that the average value associated with a species $(a,z)$ is

\begin{equation}
T_{\rm int}=\frac{\sum_q \overline{n}_{a,z,q}T_{\rm int}^{a,z,q}}{\sum_q\overline{n}_{a,z,q}}\;.
\label{eq:aveTint}
\end{equation}

\begin{figure}[tbh]
\includegraphics[width=8.5cm,angle=0]{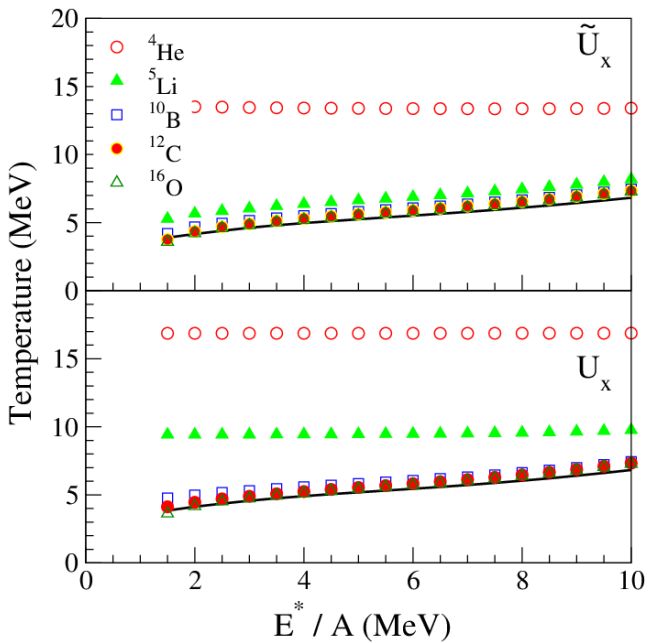}
\caption{\label{fig:tint} (Color online) Internal temperature $T_{\rm int}$, calculated from Eq.\ (\ref{eq:aveTint}), for some selected fragments produced in the breakup of a $^{120}$Nd source at different excitation energies.
In the bottom panel, the densities of states have been calculated using $U_x$ whereas the results associated with the use of $\tilde U_x$ are displayed in the top panel.
The line corresponds to the breakup temperature, given by Eq.\ (\ref{eq:tbk}). For details see the text.}
\end{figure}

\noindent
Figure\ \ref{fig:tint} displays $T_{\rm int}$ for a few selected fragments.
The source considered is the $^{120}$Nd, but we have checked that the effect of the source's size is small.
More precisely, we found that the results obtained with the $^{80}$Zr source are very similar to those shown in Fig.\ \ref{fig:tint}, small differences being noticed only for the $^{40}$Ca source.
The bottom panel of Fig.\ \ref{fig:tint} shows the results obtained with $U_x$.
One sees that, in the case of the $^4$He and $^5$Li fragments, the temperature is constant over all the excitation energy range considered, in agreement with $T_{\rm em}$ observed experimentally \cite{temperatureReviewMSU1994,temperaturesTrautmann2007,temperaturesHuang1997,ccXi,ccSerfling}, as discussed above.
However, the saturation value found in our model occurs only for the lightest fragments, depends on the species considered, and is much higher than the experimental value $T_{\rm em}\approx 5-6$ MeV.
Although the saturation observed experimentally has not been fully understood yet, its origin is clear in the framework of this model.
As may be noted from Fig.\ \ref{fig:rho}, the density of states for the $^4$He and $^5$Li nuclei has a fairly constant logarithmic derivative over a wide range of excitation energies,  and its value is equal to $1/\tau$ as seen from Eq.\ (\ref{eq:rhogc}).
It leads to $\tau=16.9$ MeV and $\tau=9.4$ MeV for the $^4$He and $^5$Li nuclei, respectively, which correspond to the values observed in Fig.\ \ref{fig:tint}.
The situation is qualitative different for the heavier fragments, since their densities of states do not exhibit the same properties as those of the $^4$He and $^5$Li nuclei, which can also be seen in Fig.\ \ref{fig:rho}.
Thus, the internal temperature of the heavier fragments follows the breakup temperature (represented by the line in Fig.\ \ref{fig:tint}) and is fairly insensitive to the species employed in the calculation.

\begin{figure}[bth]
\includegraphics[width=8.5cm,angle=0]{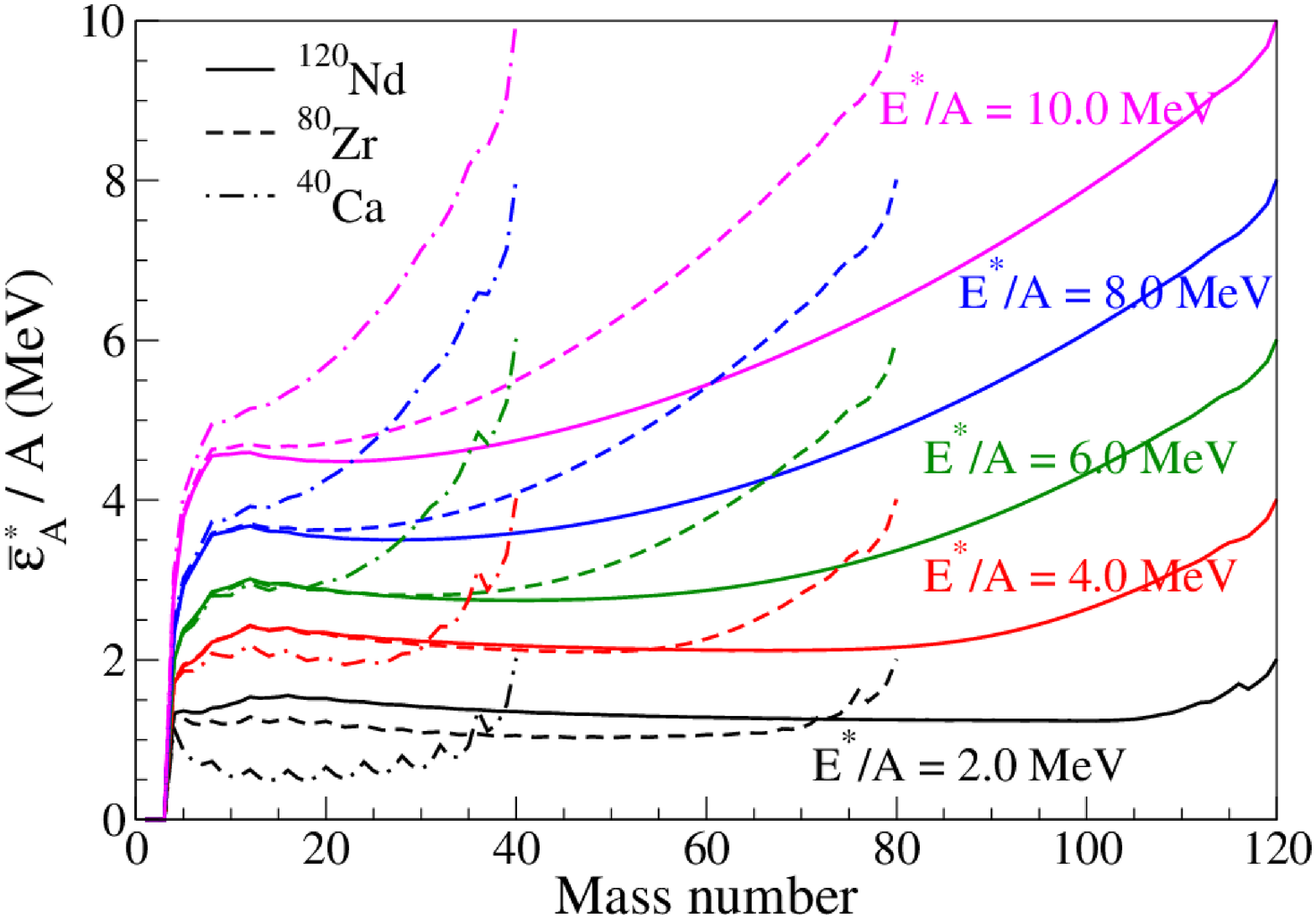}
\caption{\label{fig:eexa} (Color online) Average excitation energy per nucleon, $\overline{\epsilon}^*_A/A$, of fragments with mass number $A$, for the breakup of the sources studied in this work.
For each source, five different excitation energies are considered, {\it i. e.} $E^*/A=2.0$, $4.0$, $6.0$, $8.0$, and $10.0$ MeV.
The corresponding label is placed near the results associated with the $^{120}$Nd source, from which those related to the others may be easily located.
For details see the text.}
\end{figure}

The same conclusions hold if one uses a slightly different density of states, such as that which employs $\tilde U_x$, which reduces the region in which the constant temperature parameterization is adopted, as may be observed in Fig.\ \ref{fig:rho}.
In it, one sees that the $^4$He nucleus is the only fragment to exhibit a constant logarithmic derivative.
The corresponding internal temperatures are displayed in the top panel of Fig.\ \ref{fig:tint} for the same fragments just considered.
The temperature associated with the $^4$He is slightly lower than before because $\tau$ is smaller when one adopts $\tilde U_x$.
On the other hand, the internal temperature of the $^5$Li is significantly reduced, approaching the breakup temperature.
This is because the constant temperature region becomes much smaller, as may also be seen in Fig.\ \ref{fig:rho}.
The heavier fragments are practically not affected by $\tilde U_x$ in this context.

Our results suggest that the properties of the density of states may explain why the internal temperatures of the fragments are fairly constant over a wide range of excitation energy.
We stress that no attempt to fit the emission temperatures observed experimentally has been made.
Our aim consists in pointing out the relevance of the properties of the light fragments' state densities in the determination of their internal temperatures.
In this way, we expect that investigations on the properties of $T_{\rm int}$ may provide important information on the density of states of light nuclei.

We also remark that the standard SMM \cite{smm1,smm2} assumes that $T_{\rm int}$ is equal to the breakup and kinetic temperatures.
The present implementation suggests that this is a reasonable approximation if the constant temperature behavior of the light fragments' state densities is restricted to a very narrow excitation energy range.
Thus, the validity of this simplifying assumption relies on this qualitative property of the density of states.

We now inspect the behavior of the average excitation energy of fragments with mass number $a$ 

\begin{equation}
\overline{\epsilon}^*_a=\frac{\sum_{z,q}\overline{n}_{a,z,q}\overline{\epsilon}^*_{a,z,q}}{\sum_{z,q}\overline{n}_{a,z,q}}
\label{eq:aveexa}
\end{equation}

\noindent
and show this quantity per nucleon in Fig.\ \ref{fig:eexa}, for the different sources used in this work, at selected excitation energies values.
One sees that $\overline{\epsilon}^*_A/A$ increases with the excitation energy of the source.
For each source, $\overline{\epsilon}^*_A/A$ exhibits a plateau as a function of the mass number, whose extension diminishes as the source's excitation energy increases.
The rise of the curves at mass numbers close to the source's size is due to the contribution of the heavy remnant, which holds most of the excitation energy, tending to $E^*/A$ as the fragment's size approaches the system size.
This tendency becomes more important, leading to a faster rise, as the source's excitation energy increases.
This is due to the decrease of the largest fragment's size with the increase of $E^*/A$.
For small values of $E^*/A$, the curves exhibit a strong dependence on the source's size even for small fragments' sizes.

The sensitivity of $\overline{\epsilon}^*_A$ on the source's size is further illustrated in Fig.\ \ref{fig:ne}, which displays the distribution of the average excitation energy of the $^{16}$O fragment produced in the breakup of different sources.
At the lowest source's excitation energy, $E^*/A=2.0$ MeV, the distributions shift to higher energy values as the source sizes increase.
This picture quickly changes at $E^*/A=4.0$ MeV, where the distributions of the two heaviest sources are virtually indistinguishable, whereas that associated with the lightest source is peaked at a slightly lower value and is also narrower than the others.
The discrepancies diminish as the excitation energy increases, the distributions being very similar at $E^*/A=6.0$ MeV, although that associated with the lightest source seems to exhibit a tendency to peak at a slightly higher energy value.
This is confirmed at $E^*/A=8.0$ MeV, where its peak is clearly displaced to the right, with respect to the other distributions.
The same tendency is also observed for the $^{\rm 80}$Zr source, although with a much smaller magnitude.
These results show that the observed finite size effects are expected to be small only in a narrow excitation energy range, {\it i.e.} for $4.0 ({\rm MeV}) \lesssim E^*/A
\lesssim 7.0 ({\rm MeV})$.

\begin{figure}[bth]
\includegraphics[width=8.5cm,angle=0]{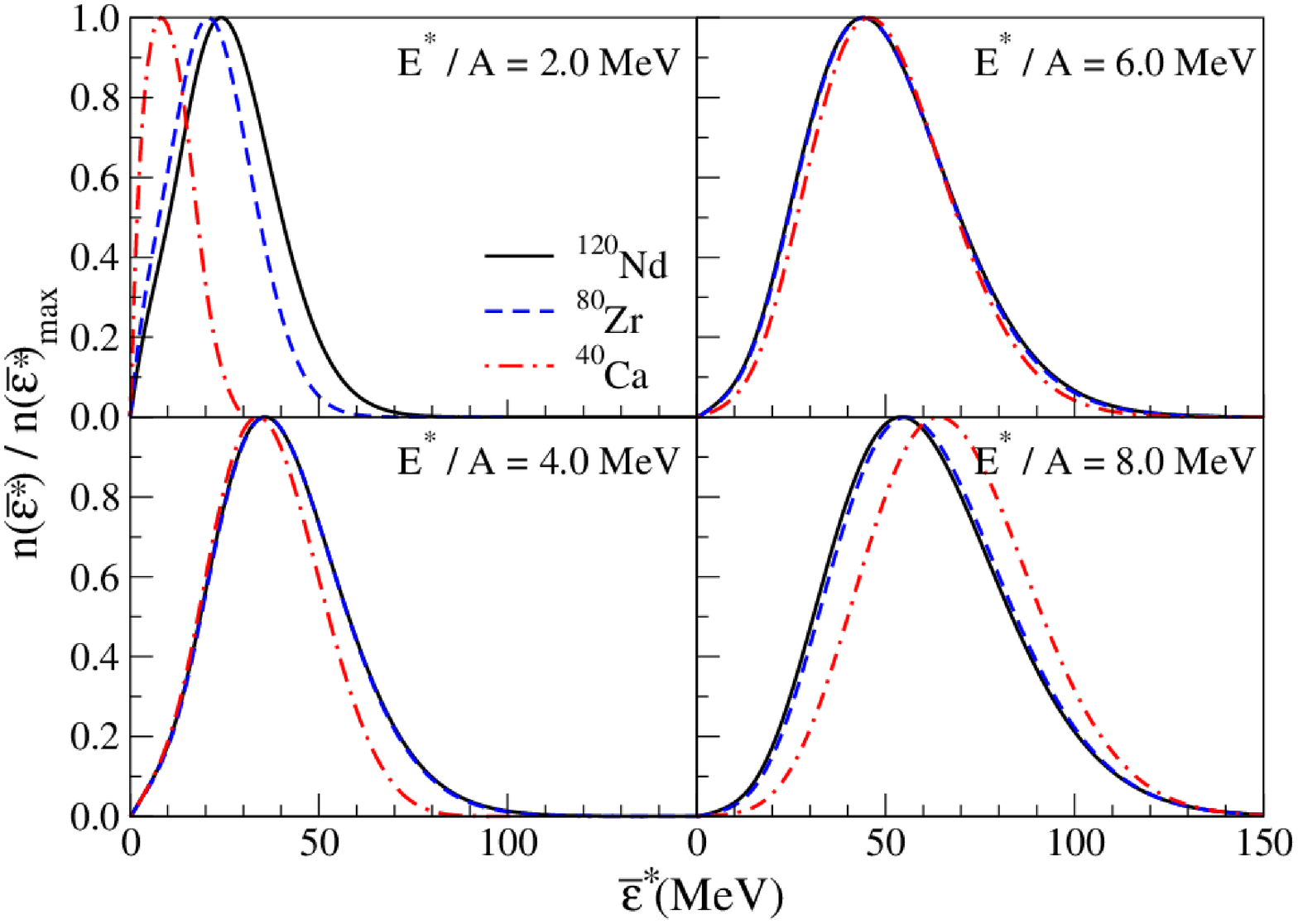}
\caption{\label{fig:ne} (Color online) Distribution of the average excitation energy of the $^{16}$O fragment produced in the breakup of the different sources used in this work at selected excitation energies.
For details see the text.}
\end{figure}

The qualitative behavior of the density of states also plays a role on the fragments' energy distribution.
This is illustrated in Fig.\ \ref{fig:eexau}, which shows $\overline{\epsilon}^*_A/A$ as a function of the fragment's mass number, obtained using both $U_x$ and $\tilde U_x$ in the density of states.
We considered the $^{120}$Nd source, but checked that similar results are obtained with the others.
One sees that a narrower constant temperature region ($\tilde U_x$) affects essentially the light fragments ($A\lesssim 15$), leading to higher excitation energy values.
The heavier fragments are much less affected, exhibiting smaller values of excitation energy due to energy conservation.
It therefore suggests that the conclusions drawn from Fig.\ \ref{fig:eexau} will not be qualitatively affected by the use of a somewhat different state density.

\begin{figure}[bth]
\includegraphics[width=8.5cm,angle=0]{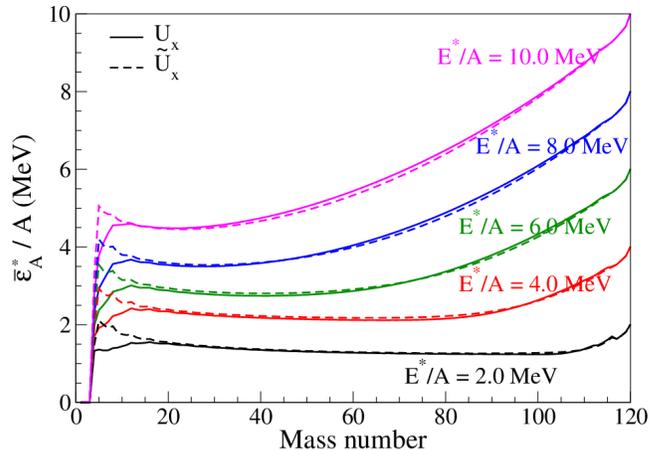}
\caption{\label{fig:eexau} (Color online) Average excitation energy per nucleon, $\overline{\epsilon}^*_A/A$, of fragments with mass number $A$, for the breakup of the $^{120}$Nd source at $E^*/A=2.0$, $4.0$, $6.0$, $8.0$, and $10.0$ MeV.
The calculations have been carried out employing the parameters $U_x$ and $\tilde U_x$ for the density of states.
For details see the text.}
\end{figure}

\end{section}

\begin{section}{Concluding Remarks}
\label{sect:conclusions}
The agreement between the fragments' internal and kinetic temperatures and the source's breakup temperature has been investigated using an implementation of the SMM which makes no a priori assumptions on the relationship among them.
It is found that the kinetic temperature is similar to the breakup temperature, except for small excitation energies.
The source's size turned out to affect the agreement, the discrepancies increasing as the size decreases.
A very good agreement in the range 2.0 MeV $\lesssim E^*/A\lesssim$ 10.0 MeV is found for source sizes $A_0\gtrsim 120$.

The internal temperatures $T_{\rm int}$ of very light fragments turned out to exhibit a constant behavior over the excitation energy range studied, 2.0 MeV $\lesssim E^*/A\lesssim$ 10.0 MeV, in qualitative agreement with the experimental observations.
However, $T_{\rm int}$ exhibits a strong sensitivity to the properties of the state densities at low excitation energies.
Our results, therefore, suggest that important information on the state density may be obtained from studies of the fragments' internal temperatures.
We also found that, as the fragments' size increases, $T_{\rm int}$ quickly approaches the breakup temperature.
The source's size is found  to play a minor role in this context.

On the other hand, it strongly influences the average excitation energy of the fragments $\overline{\epsilon}^*_A$.
It is found that $\overline{\epsilon}^*_A/A$ rises as a function of the fragments' mass number $A$,  as the system size is approached.
This effect is enhanced by the increase of the source's excitation energy.
This is due to the contribution of the heavy remnant, which holds most of the excitation energy.
However, for low excitation energy values of the source, $E^*/A\lesssim 4.0$ MeV, important differences are observed even for small fragment sizes, {\it i.e.} $A<<A_0$.
It therefore reveals that the excitation energy of the fragments is very sensitive to the source's size and this fact must be taken into account in interpreting observables related to $\overline{\epsilon}^*_A$.
The qualitative behavior of the fragments' state density at low excitation energy also turns out to affect $\overline{\epsilon}^*_A$.

In conclusion, important information on the qualitative properties of the light fragments' densities of states may be obtained from studies of the fragments' internal temperatures and from observables related to the fragments' excitation energy.
Finite size effects, associated with the source size, are also expected to influence some of the observables, such as the kinetic temperature and the fragments' excitation energy.

\end{section}

\begin{acknowledgments}
We would like to acknowledge CNPq, FAPERJ BBP grant, and FAPESP for partial financial support.
This work was supported in part by the National Science Foundation under Grant No. PHY-1102511.
We also thank the
Programa de Desarrollo de las Ciencias B\'asicas (PEDECIBA) and the
Agencia Nacional de Investigaci\'on e Innovaci\'on (ANII) for partial financial support.
\end{acknowledgments}

\bibliography{manuscript}
\bibliographystyle{apsrev4-1}

\end{document}